\documentclass{ws-procs975x65}

\newcommand{\bea}{\begin{eqnarray}}
\newcommand{\eea}{\end{eqnarray}}
\newcommand{\rxdieciocho}{\hbox{RX~J1856.5$-$3754~}}
\newcommand{\rxcerosiete}{\hbox{RX~J0720.4$-$3125~}}
\newcommand{\rbdoce}{\hbox{RBS~1223~}}
\begin{document}


\title{Joule heating in the cooling of magnetized neutron stars}

\author{Deborah N.~Aguilera$^*$ and Jos\'e A.~Pons and Juan A.~Miralles}

\address{Departamento de F\'{\i}sica Aplicada, Universidad de Alicante,\\
Apartado Correos 99\\
03080 Alicante, Spain\\
$^*$E-mail: deborah.aguilera@ua.es}

\begin{abstract}
We present 2D simulations of the cooling of neutron stars with strong magnetic fields ($B \geq 10^{13}$ G). We solve the diffusion equation in axial symmetry including the 
state of the art  microphysics that controls the cooling such
as slow/fast neutrino processes, superfluidity, as well as possible heating mechanisms. 
We study how the cooling curves depend on the the magnetic field strength and geometry. 
Special attention is given to discuss the influence of magnetic field decay. 
We show that Joule heating effects are very large and in some cases  
control the thermal evolution. 
We characterize the temperature anisotropy induced by the magnetic field 
for the early and late stages of the evolution of isolated neutron stars.
\end{abstract}


\bodymatter
\section{Introduction}\label{intro}
The observed thermal emission of neutron stars (NSs) can provide information about the matter in their interior. Comparing the theoretical cooling curves with observational data\cite{Yakovlev2004,Page2006} one can infer not only the physical conditions of the outer region (atmosphere) where the spectrum is formed but also of the poorly known interior (crust, core) where high densities are expected.

There is increasing evidence that most of nearby NSs whose thermal emission is visible in the X-ray band have a non uniform temperature distribution\cite{Zavlin2007,Haberl2007}~. There is a mismatch between the extrapolation to low energy of the fits to X-ray spectra,
and the observed Rayleigh Jeans tail in the optical band ({\it optical excess flux}), that cannot be addressed with a unique temperature (e.g. \rxdieciocho\cite{Pons2002}~, \rbdoce\cite{Schwope2007}~,  and 
\rxcerosiete\cite{Perez2006}~). 

A non uniform temperature distribution
 may be produced not only in the low density regions\cite{Greenstein1983}~, 
but also in intermediate density regions, such as the solid crust. 
Recently, it has been proposed that crustal confined magnetic fields with strengths larger than $10^{13}$ G   could be responsible for the surface thermal anisotropy \cite{Geppert2004,Azorin2006}~. 
In the crust, the magnetic field limits the movement of electrons (main responsible for the heat transport)
in the direction perpendicular to the field and the thermal conductivity 
in this direction is highly suppressed, while remains almost unaffected 
along the field lines. 

Moreover, the observational fact that most thermally emitting isolated 
NSs have magnetic fields larger than $10^{13}$ G implies that a realistic 
cooling model must include magnetic field effects. In a recent work\cite{Aguilera2007}~, first 2D simulations of the cooling of magnetized NSs have been presented. In particular, it has been stated that magnetic field decay, as a heat source, could strongly affect the thermal evolution and the observations should be reinterpreted in the light of these new results. We present the main conclusions of this work next. 

\section{Non-uniform temperature distribution induced by magnetic field}

We consider two baseline models\cite{Aguilera2007}~: Model A,  a low mass NS  with $M=1.35$~$M_{\odot}$ and Model B, a high mass NS with $M=1.63$~$M_{\odot}$. These two models correspond to the {\it minimal cooling} scenario (controlled by modified Urca neutrino emission) and the {\it fast cooling} scenario (where direct Urca operates), respectively.  
We consider crustal confined magnetic fields, keeping the geometry fixed 
and varying the magnetic field strength at the pole  ($B$). 

\begin{figure}[htb]
\centering
\psfig{file=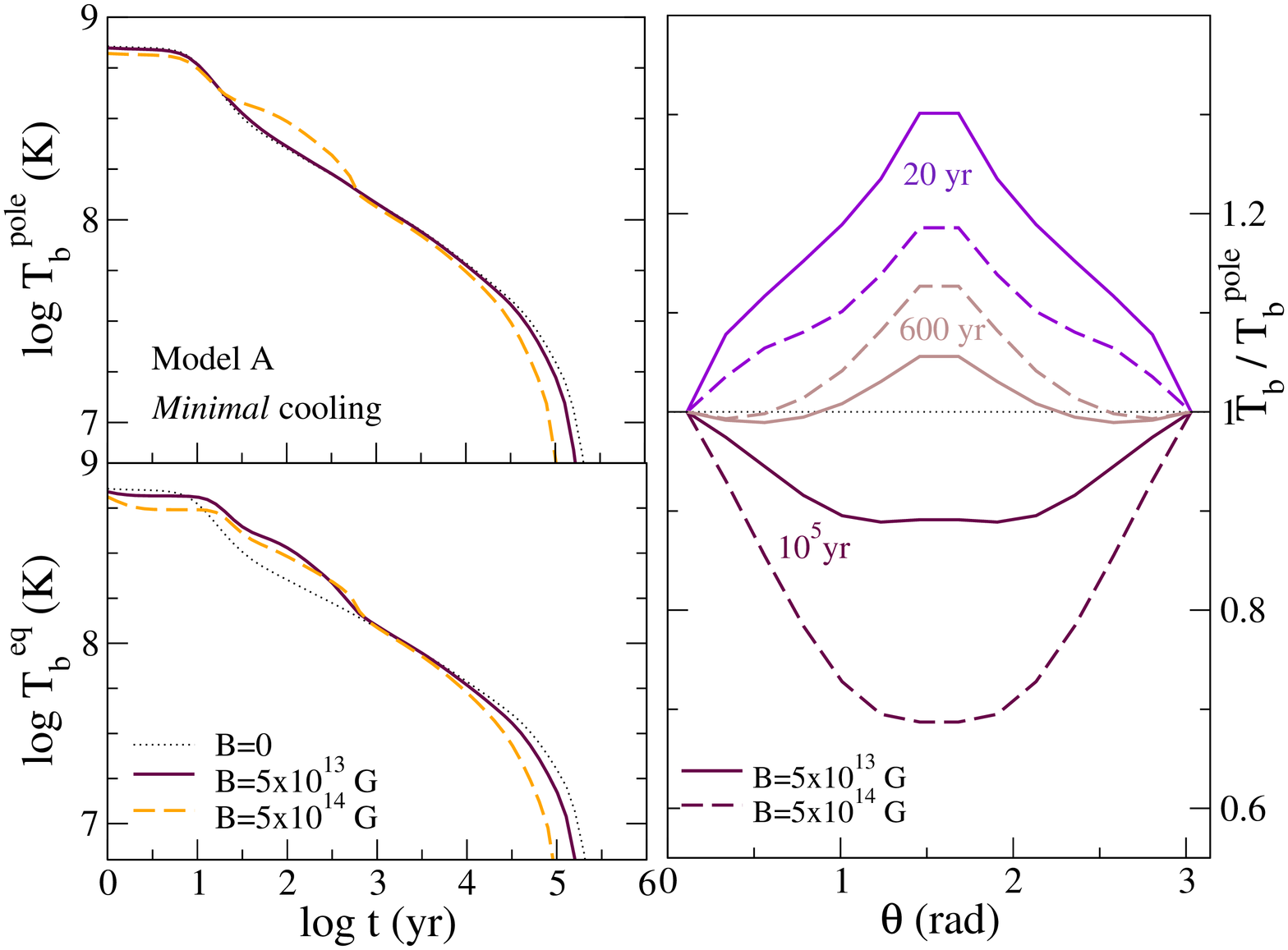,width=0.38\linewidth,angle=-90}
\psfig{file=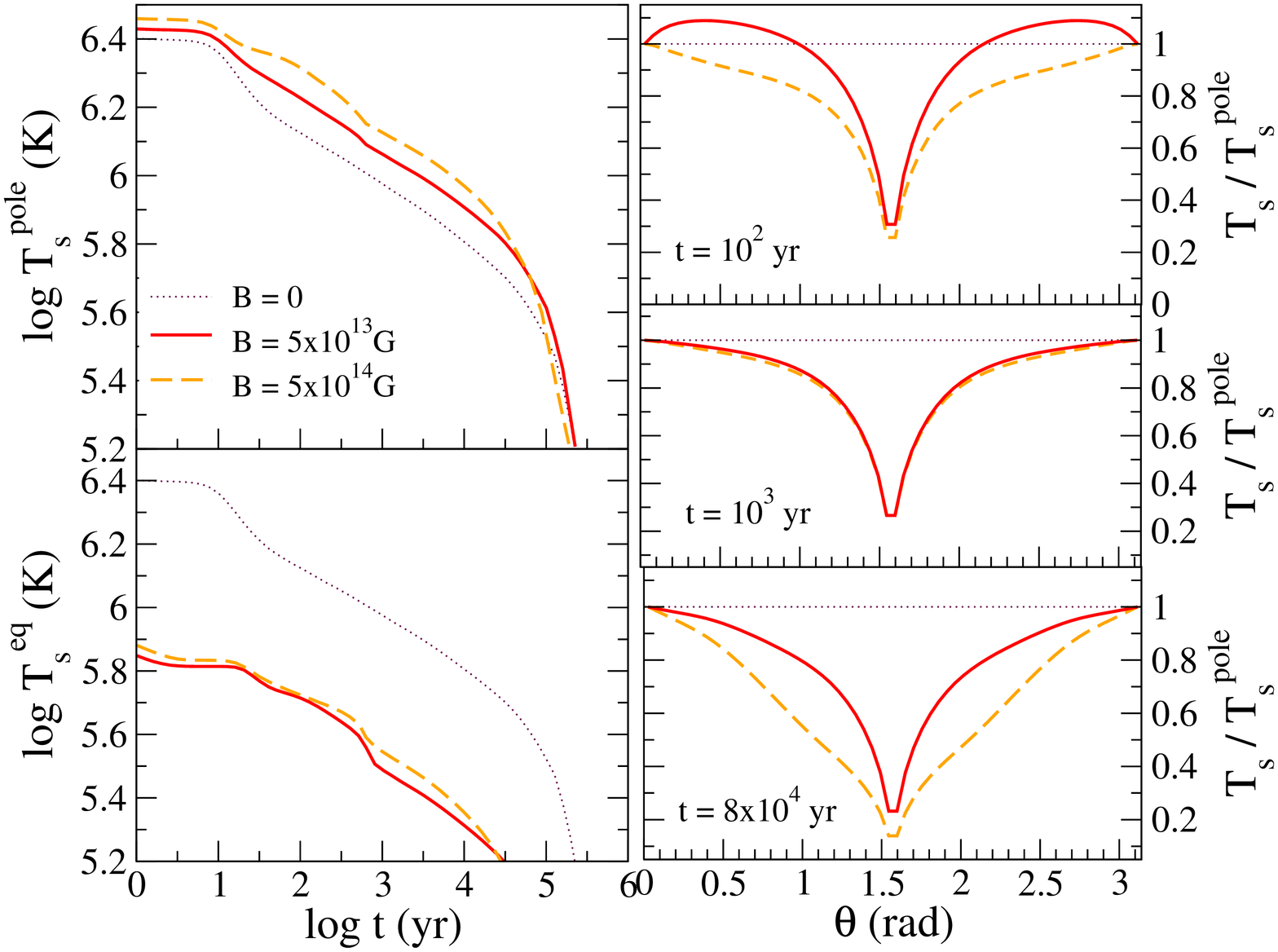,width=0.38\linewidth,angle=-90}
\caption{Cooling of strongly magnetized NSs for the Model A. 
Temperatures $T_b$ (left figure) and $T_s$ (right figure) at the pole and at the equator as a function of 
the age $t$ (left panels). Corresponding evolution of $T/T^{\rm pole}$ vs $\theta$ in the right panels.}
\label{fig_tb}
\end{figure}
For a given magnetic field,  we solve the diffusion equation in axial symmetry, considering an anisotropic thermal conductivity tensor $\hat \kappa$. The ratio of its components along and perpendicular to the field
can be defined in terms of the magnetization parameter 
 ($\omega_{B} \tau$) as $\kappa^{\parallel}_e/\kappa^{\perp}_e = 1 + (\omega_{B} \tau)^{2}$, where
$\tau$ is the electron relaxation time  and
$\omega_B $ is the electron cyclotron frequency. When $\omega_B\tau \gg 1$ the magnetic field effects
on the transport properties are crucial. 
We show the results in Fig.~\ref{fig_tb} (on the left), where the temperature at the base of the envelope $T_b$ (at $10^9$~g/cm$^3$) is shown as a function of the age $t$. We see an {\em inverted temperature distribution} with 
cooler polar caps and a  warmer equatorial belt at $t \lesssim 500$~yr for the Model A. Similar qualitative results are found for Model B.  

In Fig.~\ref{fig_tb} (on the right) we show for Model A the corresponding cooling curves but for the surface temperature $T_s$. We see that the anisotropy found at the level of $T_b$ does not 
automatically result in a similar $T_s$ distribution: the blanketing effect of the envelope overrides the inverted temperature distribution found at
intermediate ages. Thus, the equator remains always cooler than the pole, and only at early times and for strong fields we find larger surface temperatures in middle latitude regions.

\section{Magnetic field decay and Joule heating}

For the magnetic field decay, we assume the approximate solution of the diffusion equation
\bea 
B= B_0\frac{\exp{(-t/\tau_{\rm Ohm})}}{1+\frac{\tau_{\rm Ohm}}{\tau_{\rm Hall}}
(1-\exp{(-t/\tau_{\rm Ohm})})} 
\label{Btime} 
\eea 
where $\tau_{\rm Ohm}$ is the Ohmic characteristic time,  
and the typical timescale of the fast, initial stage  
is defined by $\tau_{\rm Hall}$. 
 
\begin{figure}[htb]
\begin{center}
\psfig{file=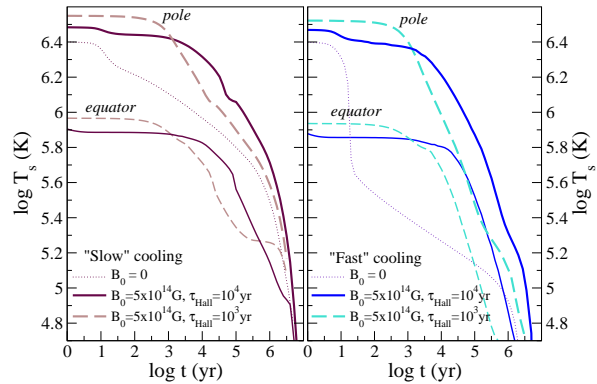,width=0.5\textwidth,angle=-90}
\end{center}
\caption{Cooling of strongly magnetized NSs with Joule heating with 
$B_0=5 \times 10^{14}$~G for Model A (left panel) and Model B (right panel). We have set $\tau_{\rm Ohm}=10^6$~yr.}
\label{fig_Joule}
\end{figure}

In the cooling curves, there is a huge effect due to the decay of such a large field: as a consequence of the heat released, $T_s$ remains much higher than in the case of constant field (Fig.~\ref{fig_Joule}). 
The strong effect of the field decay is evident for all of the pairs of parameters chosen. Notice that $T_s$ of the initial plateau is higher for shorter $\tau_{\rm Hall}$,
but the duration of this stage with nearly constant temperature is also shorter. After $t \simeq \tau_{\rm Hall}$, there is a drop in $T_s$ due to the transition from the fast Hall stage to the slower Ohmic decay.
\section{Conclusion: Towards a coupled magneto-thermal evolution}
\begin{figure}[htb]
\begin{center}
\psfig{file=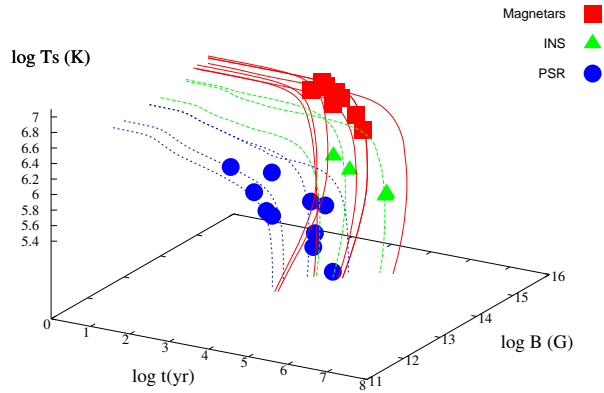,width=0.5\textwidth,angle=-90}
\end{center}
\caption{Coupled magneto-thermal evolution of isolated neutron stars\cite{Aguilera2007a}~: $T_s$ for the hot component as a function 
of $B$ and $t$. Observations: squares for magnetars ($B>10^{14}$~G), triangles for intermediate-field isolated
NSs ($10^{13}$~G$<B<10^{14}$~G) and circles for radio pulsars ($B<10^{12}$~G). Corresponding cooling curves in solid, dashed and dotted lines, respectively. }
\label{fig_coupled}
\end{figure}

The main result of this work is that, in magnetized NSs with $B> 10^{13}$~G, the decay of the magnetic field affects strongly their cooling. In particular,  there is a huge effect of 
Joule heating on the thermal evolution. In NSs born as magnetars, this effect plays a key role in maintaining them warm for a long time. Moreover,  it can also be important in  
high magnetic field radio pulsars and in radio--quiet isolated NSs. As a conclusion, the thermal and magnetic field evolution of a NS is at least a two parameter space (Fig~\ref{fig_coupled}), and a first step towards a coupled magneto-thermal evolution has been given in this work.  

\end{document}